\documentclass[sigconf]{acmart}

\usepackage[ruled,linesnumbered,vlined]{algorithm2e}
\usepackage{amsmath}
\usepackage[normalem]{ulem}
\usepackage{pifont}
\usepackage{makecell}
\usepackage{listings}
\usepackage{minted}
\usepackage{subcaption}
\usepackage{float}
\usepackage{tikz}
\usepackage{multirow}
\usepackage{tabularray}
\usepackage{tabularx}
\usepackage{stackengine}
\usepackage[export]{adjustbox}
\usepackage{xcolor} 
\usepackage{soulutf8}
\usepackage{xspace}
\usepackage{graphicx}
\usepackage{booktabs}
\usepackage{etoolbox}
\usepackage{lipsum} 
\usepackage{enumitem}
\usepackage{wrapfig}
\usepackage{hyperref}

\newcommand{\cmt}{\texttt{Cement2}\xspace}
\newcommand{\frontend}{\texttt{CMT2-rs}\xspace}
\newcommand{\ctir}{\texttt{CTIR}\xspace}

\newcommand{\x}{\texttimes\xspace}

\definecolor{lightgray}{rgb}{0.98, 0.98, 0.98}
\definecolor{brickred}{rgb}{0.8, 0.25, 0.33}
\definecolor{darkspringgreen}{rgb}{0.09, 0.45, 0.27}
\definecolor{purple}{rgb}{0.39, 0.06, 0.80} 

\definecolor{comment}{RGB}{0, 128, 0}        
\definecolor{keyword}{RGB}{0, 0, 255}        
\definecolor{string}{RGB}{163, 21, 21}       
\definecolor{number}{RGB}{9, 134, 88}        
\definecolor{type}{RGB}{38, 127, 153}        
\definecolor{function}{RGB}{121, 94, 38}     
\definecolor{variable}{RGB}{0, 16, 128}      
\definecolor{operator}{RGB}{0, 0, 0}         

\newcommand{\dhalfcheck}{\textcolor{darkspringgreen}{\ding{51}}\textsuperscript{\textcolor{brickred}{\kern-0.5em\tiny\ding{55}}}}
\newcommand{\dcheck}{\textcolor{darkspringgreen}{\ding{51}}}
\newcommand{\dx}{\textcolor{brickred}{\ding{55}}}

\newcommand{\uv}[1]{{\color{blue} #1}}

\definecolor{LightGray}{gray}{0.97}

\newcommand{\compactparagraph}[1]{{\noindent\emph{\textbf{#1.}}}\xspace}

\newcolumntype{s}{>{\hsize=.5\hsize}X}

\makeatletter
\patchcmd{\minted@colorbg}{\medskip}{}{}{}
\patchcmd{\endminted@colorbg}{\medskip}{}{}{}
\makeatother

\AtBeginDocument{%
  }

\setcopyright{none} 
\renewcommand\footnotetextcopyrightpermission[1]{} 
\author{Youwei Xiao}
\affiliation{%
  \institution{Peking University}
  \country{China}}
  
\author{Zizhang Luo}
\affiliation{%
  \institution{Peking University}
  \country{China}}

\author{Weijie Peng}
\affiliation{%
  \institution{Peking University}
  \country{China}}

\author{Yuyang Zou}
\affiliation{%
  \institution{Peking University}
  \country{China}}

\author{Yun Liang}
\affiliation{%
  \institution{Peking University}
  \country{China}}

\begin{document}

\title{\cmt: Temporal Hardware Transactions for High-Level and Efficient FPGA Programming}


\begin{abstract}

Hardware design faces a fundamental challenge: raising abstraction to improve productivity while maintaining control over low-level details like cycle accuracy. Traditional RTL design in languages like SystemVerilog composes modules through wiring-style connections that provide weak guarantees for behavioral correctness. While high-level synthesis (HLS) and emerging abstractions attempt to address this, they either introduce unpredictable overhead or restrict design generality. Although transactional HDLs provide a promising foundation by lifting design abstraction to atomic and composable rules, they solely model intra-cycle behavior and do not reflect the native temporal design characteristics, hindering applicability and productivity for FPGA programming scenarios.

We propose temporal hardware transactions, a new abstraction that brings cycle-level timing awareness to designers at the transactional language level. Our approach models temporal relationships between rules and supports the description of rules whose actions span multiple clock cycles, providing intuitive abstraction to describe multi-cycle architectural behavior. We implement this in \cmt, a transactional HDL embedded in Rust, enabling programming hardware constructors to build both intra-cycle and temporal transactions. \cmt's synthesis framework lowers description abstraction through multiple analysis and optimization phases, generating efficient hardware. With \cmt's abstraction, we program a RISC-V soft-core processor, custom CPU instructions, linear algebra kernels, and systolic array accelerators, leveraging the high-level abstraction for boosted productivity. Evaluation shows that \cmt does not sacrifice performance and resources compared to hand-coded RTL designs, demonstrating the high applicability for general FPGA design tasks.
\end{abstract}


\maketitle

\section{Introduction}
\label{sec:introduction}

Hardware design at the register transfer level (RTL) is becoming increasingly challenging as architectural innovations demand more complex implementations~\citep{taylor_basejump_2018,amd_inc_vitis_2025-1,santos_scalable_2022}. Traditional RTL design in languages like SystemVerilog and VHDL composes larger modules by connecting ports of smaller ones, but this wiring-style composition provides weak guarantees. Behavioral correctness cannot be guaranteed, causing data communication errors like producer-consumer mismatches~\citep{ma_debugging_2022}. For FPGA programming, this fundamental limitation causes a significant gap between architecture design and hardware implementation. Designers often need to describe their architecture design in two models: higher-level simulators~\citep{lowe-power_gem5_2020,sanchez_zsim_2013} for fast idea validation, and tedious RTL design for detailed implementation on FPGAs. The potentially inaccurate simulation results and missing area and power information can lead to wrong architectural decisions and impede iteration speed. Besides, designers must invest significant cognitive load and development cost to ensure correct implementation aligned with design intent.
 
Raising hardware design abstraction is necessary to address these challenges. Describing hardware at a higher level can avoid the tedious and error-prone composition. However, choosing the appropriate abstraction level presents a non-trivial tradeoff. High-level synthesis (HLS)~\citep{amd_inc_vitis_2025,josipovic_dynamically_2018,google_inc_xls_2025} takes untimed software descriptions and generates hardware designs, boosting productivity. However, it often produces unpredictable performance and resource overheads~\citep{nigam_predictable_2020}, since it loses control over low-level details, such as cycle accuracy and resource overheads, limiting its applicability. 
Accelerator design languages~\citep{nigam_compiler_2021,kim_unifying_2023,majumder_hir_2024,xu_hector_2022,nigam_predictable_2020} are also too specific. Other emerging FPGA programming approaches, including Cement~\cite{xiao_cement_2024} and others~\citep{skarman_spade_2022, zagieboylo_pdl_2022,nigam_modular_2023,durst_type-directed_2020}, add latency-sensitive/-insensitive information but put more constraints on design architecture and hardware composition manners.

An ideal FPGA programming methodology should raise the design abstraction level while maintaining control over clock-cycle timing and low-level hardware details. Transactional HDLs~\citep{nikhil_bluespec_2004,choi_kami_2017,bourgeat_essence_2020} provide a promising foundation by abstracting hardware design as behavioral rules, composable logic units with execution atomicity. However, prior rule-based works are restricted to intra-cycle logic and cannot reflect hardware's temporal behavior or architecture design intent. In practice, designers either adopt a latency-insensitive design style to eliminate temporal concerns with extra hardware overheads or manually coordinate intra-cycle rules for efficient latency-sensitive design at the expense of productivity.

We propose a new abstraction, \textit{temporal hardware transactions}, which brings cycle-level timing awareness to designers at the language level, for high-level FPGA programming. We model temporal relationships between rules, building an intuitive temporal view of rules' execution across cycles. To further boost productivity in describing multi-cycle behaviors, we introduce multi-cycle rules whose actions span multiple clock cycles under specified constraints. Our abstraction's temporal behavior modeling enables comprehensive inter-cycle hardware analysis, checking, and optimizations, leading to compiler-enforced correctness and efficiency. Moreover, it employs a multi-phase synthesis flow to transition from high-level abstraction to low-level, while avoiding the introduction of unnecessary overhead, resulting in efficient hardware implementation.
Both (micro)architecture and hardware design phases of FPGA programming benefit from temporal hardware transactions. Architects need a precise estimation of performance and hardware overheads to guide (micro-) architectural optimization. With our abstractions, architects can accurately model and implement hardware using a single behavioral interface, rather than multiple models across behavioral and structural levels~\cite{lowe-power_gem5_2020,sanchez_zsim_2013,lockhart_pymtl_2014}. This greatly improves the development productivity. For hardware designers, our abstraction provides a more productive design methodology. It allows behavior description with temporal intuition and enables the compiler to conduct rich behavioral checking for early error detection. The powerful synthesis flow not only handles hybrid latency-sensitive/-insensitive but also performs retiming on multi-cycle rules, generating an efficient RTL implementation. Specifically, for HLS designers who care about performance, our abstraction provides more accurate control over details while retaining synthesis features; for RTL designers, the abstraction helps to boost productivity. Moreover, the compiler can detect many design errors early, greatly reducing the cost and effort required for fixing them later (e.g., manually writing test benches, debugging, etc.)



We implement the novel abstraction in the \cmt framework, which is abbreviated as \texttt{CMT2}. The frontend language \frontend is a transactional HDL embedded in Rust that enables creating hardware constructors to describe hardware transactions with temporal behavior modeling. \cmt represents constructed rules in \ctir (\cmt Transaction Intermediate Representation), providing a unified representation for both intra-cycle and temporal hardware transactions. The synthesis framework conducts temporal scheduling, temporal partitioning, and temporal implementation, generating a high-quality RTL implementation for FPGA deployment.

Our contributions are:
\begin{itemize}[leftmargin=*]
\item The novel temporal hardware transactions that raise abstraction while maintaining control over FPGA programming details;
\item The \uv{open-source} \cmt framework including a Rust frontend and \ctir implementing temporal hardware transactions;
\item A synthesis flow that efficiently generates high-quality RTL implementation for FPGA deployment.
\end{itemize}

We evaluate \cmt through four case studies on FPGA: building a RISC-V soft core, extending the core with custom instructions, describing linear algebra kernels, and designing systolic array accelerators. Experiments show that \cmt's soft core achieves a higher frequency at 377MHz with lower resource usage compared to the Sodor~\citep{noauthor_ucb-barriscv-sodor_2025} core designed in Chisel~\citep{bachrach_chisel_2012}. For other case studies, \cmt boosts design productivity and achieves comparable or better performance and hardware quality than human-crafted RTL designs. The evaluation demonstrates that temporal hardware transactions facilitate various design scenarios, and \cmt is a productive solution for general FPGA programming tasks. 
\section{Background and Motivation}

\label{sec:background}

This section discusses abstraction levels available for FPGA programming and analyzes their tradeoffs, as summarized in \autoref{tab:compare}.

\begin{figure}[t]
  \centering
  \begin{subfigure}[b]{\linewidth}
    \centering
    \includegraphics[width=\linewidth]{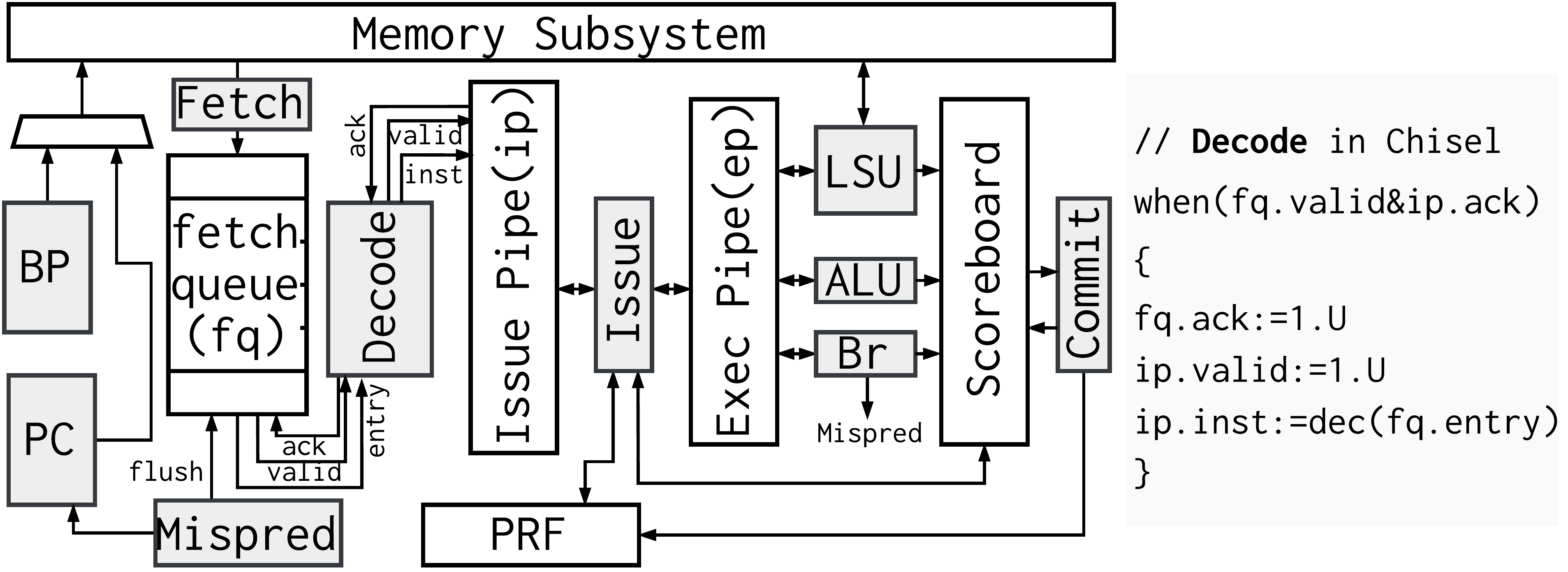}
    \caption{RTL}
    \label{fig:cpu-rtl}
  \end{subfigure}
  \\
  \hfill
  \begin{subfigure}[b]{.37\linewidth}
    \centering
    \includegraphics[width=\linewidth]{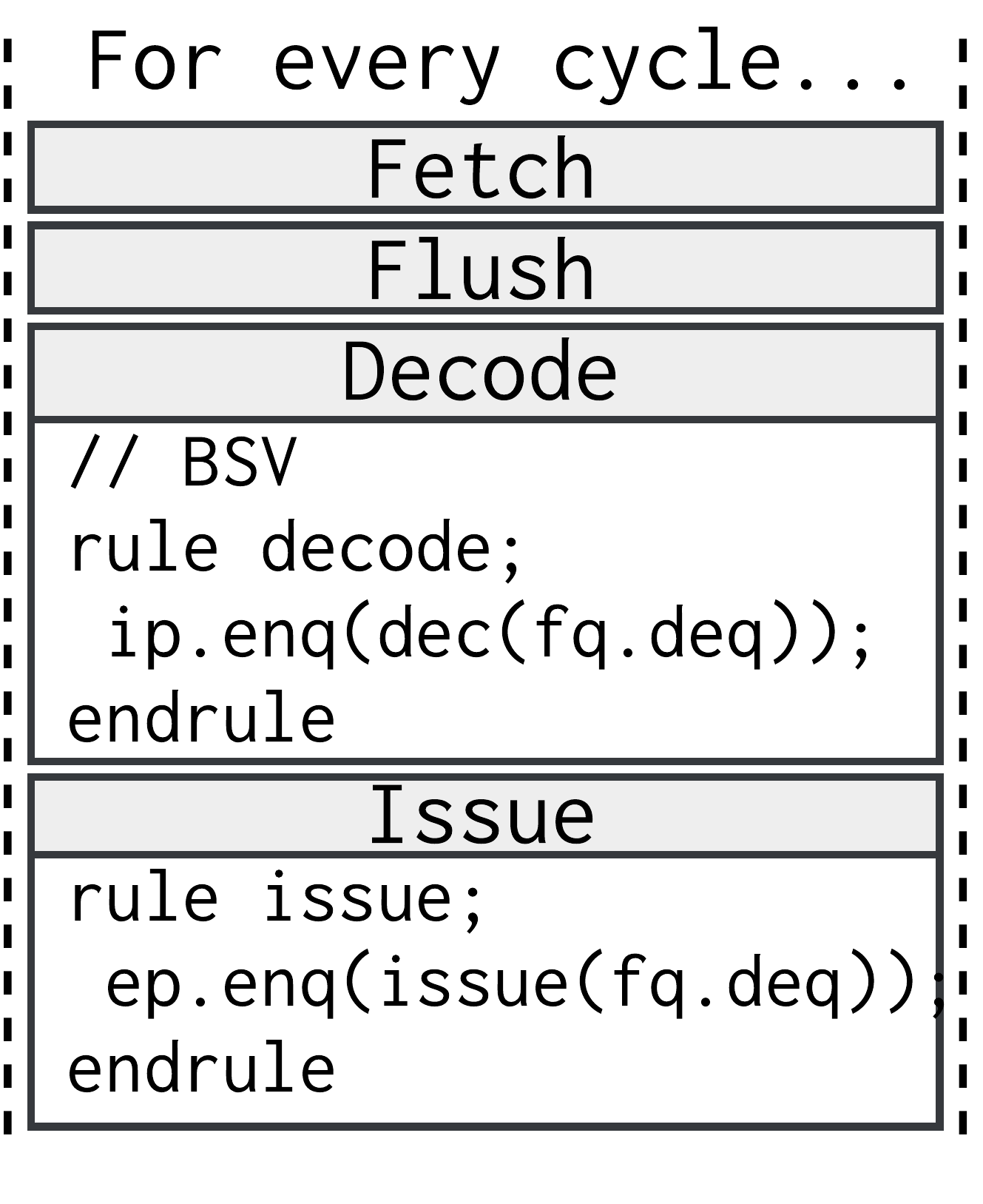}
    \caption{Transaction}
    \label{fig:cpu-rule}
  \end{subfigure}
  \hfill
  \begin{subfigure}[b]{.62\linewidth}
    \centering
    \includegraphics[width=\linewidth]{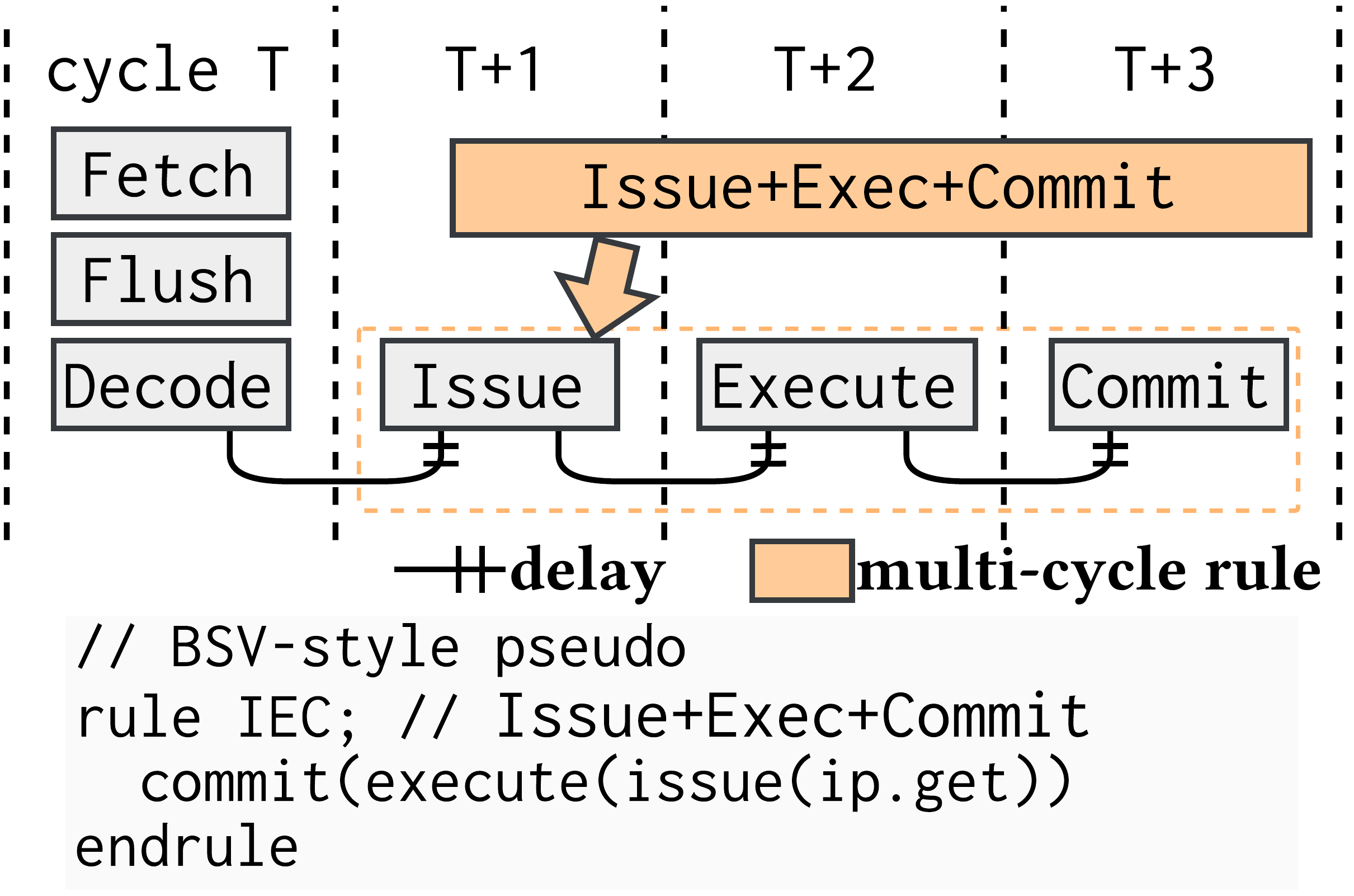}
    \caption{Temporal transaction}
    \label{fig:cpu-temporal}
  \end{subfigure}
  \hfill
  \caption{{Motivating example: illustrating the 5-stage CPU core pipeline described in different abstraction levels.}}
  \label{fig:cpu-motivate}
\end{figure}


\begin{table*}[t]
\small
\caption{Comparison of hardware and architecture design approaches. The table presents: (1) composition description manner, (2) inter-cycle behavior description support, (3) low-level control over details ("Cycle" for clock-cycle timing, and "Register" for register instantiation and access), (4) hardware overheads (performance and resources), and (5) design generality. }
\setlength{\tabcolsep}{2pt}
\renewcommand{\arraystretch}{0.95}
\centering
\begin{tabular}{|l|c|c|cc|ll|c|}
\hline
\multirow{2}{*}{\textbf{Approaches}} & \multirow{2}{*}{\textbf{Composition}} & \multirow{2}{*}{\begin{tabular}[c]{@{}l@{}}\textbf{Inter-cycle}\\ \textbf{ behavior}\end{tabular}} & \multicolumn{2}{c|}{\textbf{Control over...}} & \multicolumn{2}{c|}{\textbf{Hardware overhead...}} & \multirow{2}{*}{\textbf{Generality}}  \\
 & & & \textbf{Cycle} & \textbf{Register} & \textbf{Performance} & \textbf{Resources} & \\
\hline
RTL~\citep{bachrach_chisel_2012,design_automation_standards_committee_ieee_2024}       & \textcolor{brickred}{Wiring}    & \dx       & \dcheck & \dcheck & \qquad \textcolor{darkspringgreen}{Low}    & \qquad \textcolor{darkspringgreen}{Low}    & \dcheck \\
Calyx~\citep{nigam_compiler_2021}       & \textcolor{brickred}{Wiring}    & \dcheck       & \dcheck & \dcheck & \qquad \textcolor{darkspringgreen}{Low}  & \qquad \textcolor{darkspringgreen}{Low}   & \dhalfcheck\\
Filament~\citep{nigam_modular_2023}  & \textcolor{brickred}{Wiring}    & \dcheck   & \dcheck & \dcheck & \qquad \textcolor{darkspringgreen}{Low}    & \qquad \textcolor{darkspringgreen}{Low}    & \dx \\
Cement~\citep{xiao_cement_2024}  & \textcolor{brickred}{Wiring}    & \dcheck   & \dcheck & \dcheck & \qquad \textcolor{darkspringgreen}{Low}    & \qquad \textcolor{darkspringgreen}{Low}    & \dhalfcheck \\
HLS~(static~\citep{amd_inc_vitis_2025}, dynamic~\citep{josipovic_dynamically_2018})      & \textcolor{darkspringgreen}{Call}      & \dcheck   & \dx     & \dx     & \qquad \textcolor{brickred}{High}   & \qquad \textcolor{brickred}{High} & \dx \\
PDL~\citep{zagieboylo_pdl_2022}       & \textcolor{darkspringgreen}{Call}      & \dcheck   & \dcheck & \dx     & \qquad \textcolor{brickred}{High}  & \qquad \textcolor{brickred}{High}  & \dx \\
Transaction~\citep{nikhil_bluespec_2004,bourgeat_essence_2020}      & \textcolor{darkspringgreen}{Call}      & \dx       & \dcheck & \dcheck & \qquad \textcolor{darkspringgreen}{Low}    & \qquad \textcolor{darkspringgreen}{Low}   & \dcheck \\
{Simulation models}~\citep{lowe-power_gem5_2020,sanchez_zsim_2013}     & \textcolor{darkspringgreen}{Call}      & \dcheck   & \dhalfcheck & \dx & \qquad \textcolor{brickred}{Unavailable}  & \qquad \textcolor{brickred}{Unavailable} & \dcheck \\
\hline
\cmt (this work)     & \textcolor{darkspringgreen}{Call}      & \dcheck   & \dcheck & \dcheck & \qquad \textcolor{darkspringgreen}{Low}    & \qquad \textcolor{darkspringgreen}{Low}   & \dcheck \\
\hline
\end{tabular}

\label{tab:compare}
\end{table*} 

\subsection{Hardware Abstractions and Tradeoffs}
\label{sec:hardware-abstractions}

Traditional hardware design at the register-transfer level (RTL)~\citep{design_automation_standards_committee_ieee_2024,bachrach_chisel_2012} provides a structural description of hardware components. In RTL, data flows between registers, and logical operations are described through instantiation and structural wires. For example, in a processor pipeline as shown in \autoref{fig:cpu-rtl}, pipeline stages are divided by register-based stateful components like \texttt{fetch queue} and \texttt{issue pipeline register}, whose ports are connected by combinational logic blocks like \texttt{Decode} and \texttt{Issue}. {The code block on the right presents Chisel~\citep{bachrach_chisel_2012} description of \texttt{Decode}, that structurally connects all wires from \texttt{fetch queue} and \texttt{issue pipeline register}.} While this low-level description gives designers precise control over details, it has significant drawbacks: \textit{structural composition requires designers to manually operate wires without behavior promises provided by the language abstraction}, making large designs verbose, error-prone, and hard to debug during RTL simulation~\cite{veripool_verilator_2025} or FPGA running~\cite{zuo_vidi_2023}.

{Although veteran RTL designers can avoid certain language pitfalls with the help of lint tools~\citep{synopsys_vc_2025}, they still can make incorrect designs. For instance, \texttt{fetch queue} is structurally connected to logic blocks including \texttt{Fetch}, \texttt{Decode}, and \texttt{Mispred}, with latency-insensitive protocol signals (\texttt{valid}s and \texttt{ack}s) exposed.  One potential design mistake is that \texttt{fetch queue}'s \texttt{ack} signal for \texttt{Fetch} is not disabled by the high \texttt{flush} signal. When a misprediction is detected, the \texttt{Mispred} block will raise the \texttt{flush} signal to discard existing instructions in \texttt{fetch queue}, while the \texttt{Fetch} block may still fetch a new instruction from the wrong branch and enqueue it to \texttt{fetch queue} at the same cycle, causing flush failure and execution of wrong instructions. However, such errors cannot be checked and located by RTL tools. The reason is that \textit{RTL description does not model hardware in the behavioral manner and is unaware of conflicts among behaviors that manipulate shared states.} From a behavioral view, both \texttt{Mispred}'s \textit{flush} action and \texttt{Fetch}'s \textit{enque} action try to update the \texttt{fetch queue}, and the \textit{flush} action should prevent the \textit{enque} action at the same cycle. Similar issues are common among CPU pipeline components, such as \texttt{Scoreboard}~\citep{zhang_composable_2018}}.

To overcome RTL's drawbacks, various approaches have been proposed to raise the abstraction level of hardware design. As shown in \autoref{tab:compare}, high-level synthesis (HLS)~\citep{amd_inc_vitis_2025,josipovic_dynamically_2018} takes untimed software descriptions where modules are composed through function calls. While this provides the highest abstraction level and productivity, it often produces unpredictable performance since it highly relies on heuristics and lacks detailed timing control. For example, HLS cannot synthesize a CPU pipeline with data forwarding and branch prediction since those features cause unresolved inter-iteration dependencies in the source software loop. Other approaches target specific hardware design patterns. PDL~\citep{zagieboylo_pdl_2022} generates pipelines from behavioral descriptions with CPU-specific features, but it cannot describe general out-of-order execution. Calyx~\citep{nigam_compiler_2021} and Filament~\citep{nigam_modular_2023} generate efficient multi-cycle or pipelined designs from structural descriptions enhanced with control flow language or timeline type system, but they only target specific accelerator designs and cannot model the pipelined processor's complex features. Cement~\cite{xiao_cement_2024} also combines RTL description and control flow language, providing control over cycle accuracy for deterministic FPGA programming. Notably, both Calyx and Cement can be used without their high-level features, falling back to RTL design. Besides, they still suffer from RTL's error-prone features and cannot detect conflicts among behaviors that manipulate shared states. Overall, none of the approaches support general design tasks due to their over-specialized abstractions.

\subsection{Transactional Hardware Design}
\label{sec:transactional-hdl}

Transactional hardware design approaches~\citep{nikhil_bluespec_2004,bourgeat_essence_2020,choi_kami_2017} organize hardware as collections of atomic transactions. We provide a formal definition of a transactional module:

\begin{definition}[Transactional Hardware Module]
\label{def:transaction-hm}
A transactional hardware module is denoted as $M = \langle I, R, S \rangle$, where $I$ is a set of instances, $R$ is a set of rules, and $S$ is a set of binary relations for scheduling priority. Each rule $\langle id, g, f = a~|~\lambda x. a \rangle \in R$ is defined as a guarded atomic action. Here, $g$ is the rule guard, a \textit{side-effect-free} boolean predicate for explicit rule fire conditions, and $f$ is the rule fire logic. There are two types of rules: \textit{always} ($f=a$) executes proactively, and \textit{method} ($f=\lambda x. a$) executes when called with arguments $x$ and returns values. Action $a$ is a set of expressions and method calls, including register read and write. Each precedence relation ($\prec$) $\langle r_i, r_j \rangle \in S$ specifies that the rule $r_i$ must execute earlier than rule $r_j$ in every clock cycle.
\end{definition}

\autoref{fig:cpu-rule} illustrates the CPU pipeline described as hardware transactions and presents the BSV~\citep{nikhil_bluespec_2004} description, for example. Specifically, \texttt{Decode} and \texttt{Issue} logic are described as two \emph{always} rules, both of which operate stateful components, such as dequeueing \texttt{fetch queue}, and enqueueing \texttt{issue pipeline register}. In this way, hardware transactions provide a \textit{behavioral} description. 

Transactional HDLs' execution model guarantees \textit{guarded atomicity} of rules: \textit{one rule can execute only when its guard holds and all method calls in the fire logic can execute, and they are executed atomically}. Besides, it resolves \textit{conflicts} between rule behaviors that manipulate shared states. Specifically, for module $M$, it generates a \textit{scheduler} according to the precedence relation set $S$ with a total order to execute the rules in every cycle. For example, we assume that \texttt{fetch queue}'s rules have precedence relations: \texttt{deq}$\prec$\texttt{enq}$\prec$\texttt{flush}, then for a reasonable CPU pipeline schedule order [\texttt{Flush}, \texttt{Decode}, \texttt{Fetch}], the execution of \texttt{Flush}, which calls \texttt{flush}, at an earlier place will block both \texttt{Decode} and \texttt{Fetch} due to the \textit{precedence violations} of \texttt{deq}$\prec$\texttt{flush} and \texttt{enq}$\prec$\texttt{flush}. This mechanism helps hardware designers avoid the design mistakes discussed in \autoref{sec:hardware-abstractions}.

However, current transactional HDLs are limited by their intra-cycle semantics. They lack a temporal view, as all rules stand side by side and try to fire within every clock cycle. This causes the following drawbacks: (1) tedious human efforts for coordinating multi-cycle behaviors, such as describing individual rules for every logic block in \autoref{fig:cpu-rule} and composing them into a complete pipeline by describing rules' manipulation on pre-instantiated stateful components, (2) inability of analysis, checking and optimization for inter-cycle hardware behaviors, such as latency sensitivity, (3) limited synthesis capabilities, such as forcing designers to manually split a rule on critical path into multiple rules to meet timing-closure constraints through a trial-and-error process, and (4) anti-intuitive design experience, showing no pipeline stages with temporal relationships in \autoref{fig:cpu-rule}, misaligned with human intuition.

\subsection{Motivation}
\label{sec:motivation}

The fundamental challenge in advanced FPGA programming is striking the right balance between abstraction and control. We also analyze features of architecture simulators~\citep{lowe-power_gem5_2020,sanchez_zsim_2013}, identifying key insights to overcome the drawbacks of transactional HDLs. \autoref{tab:compare} shows that detailed simulation models, such as gem5 O3CPU~\citep{gem5org_gem5_nodate}, retain control over cycle and support general design. The key to this success is that they provide flexible, pure-software manipulation of temporal behavior. For example, the gem5 O3CPU model provides a \texttt{IEW} stage to handle the behavior of dispatching, issuing, executing, and writing back instructions. However, such temporal behavior across multiple clock cycles cannot be described in existing transactional HDLs. This motivates us to provide explicit temporal modeling support to hardware transactions. Accordingly, we propose \textit{temporal hardware transactions}, which intuitively describe pipeline stages and their temporal relationships and support joint multi-cycle behavioral description of multiple stages, as shown in \autoref{fig:cpu-temporal}. {The pseudo-description of multiple stages as one rule \texttt{IEC} shows that our abstraction eliminates the tedious human efforts of describing multiple individual rules that coordinate through manipulating shared instances, as did in \autoref{fig:cpu-rule}.}

Moreover, implementation overheads must be carefully considered for hardware design, which are beyond the concern of simulation models. We need synthesis techniques to translate between neighboring abstraction levels while preserving implementation efficiency. We implement the insights above in the \cmt framework, which addresses the fundamental tradeoff between design productivity and hardware quality. \autoref{tab:compare} shows that our approach solves the intra-cycle limitations of prior transactional HDLs and uses fewer hardware overheads on the CPU pipeline example, demonstrating \textit{temporal hardware transactions} as a promising high-level abstraction for hardware and architecture design.

\section{Frontend and Core Abstraction}
\label{sec:lang-abstraction}

In this section, we introduce the frontend language of our methodology, \frontend, which uniformly supports both intra-cycle transactions and \textit{temporal hardware transactions}.

\subsection{\frontend}
\label{sec:lang-frontend}

\begin{figure*}[t]
  \centering
  \includegraphics[width=\linewidth]{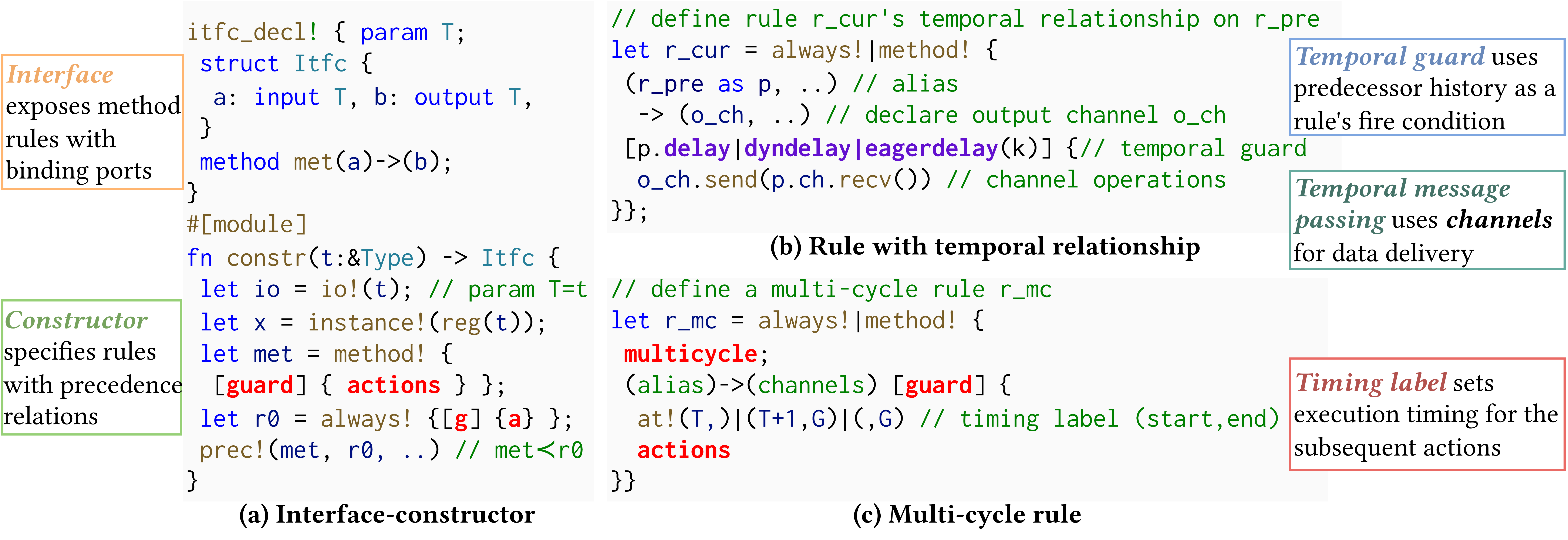}
  \caption{{Syntax of \frontend provides a unified description for intra-cycle and temporal hardware transactions.}}
  \label{fig:syntax}
\end{figure*}

\frontend is a modern transactional HDL embedded in Rust. It implements \autoref{def:transaction-hm} in an \textit{interface-constructor} pattern, whose basic syntax is shown in \autoref{fig:syntax}a. {An \textit{interface}, declared by an \texttt{itfc\_decl!} block, declares the \textit{method} rules to be called. Specifically, the \texttt{param T} defines a data type parameter, and the \texttt{struct} definition specifies the ports with directions and data types. \autoref{fig:syntax}a defines an interface named \texttt{Itfc}, which has two ports: input \texttt{a} and output \texttt{b}, both of which have the parametric data type \texttt{T}. It declares a method named \texttt{met}, which takes an argument through port \texttt{a} and returns a result through port \texttt{b}. A \textit{constructor}, defined by a \texttt{\#[module]} function, fills instances and rule implementations to construct a module of the given interface. \autoref{fig:syntax}a defines a constructor named \texttt{constr}. It builds a module of the interface \texttt{Itfc} given a specific data type \texttt{t}. The \texttt{io!} statement specifies the data type parameter \texttt{T} as \texttt{t}, and the returned variable \texttt{io} can be used to access ports (e.g., \texttt{io.a}). The \texttt{instance!} statement instantiates a register by calling the \texttt{reg} constructor. The \texttt{method!} and \texttt{always!} statements define \textit{method} and \textit{always} rules, respectively, where the \textit{guard} expression is enclosed by brackets and the fire \textit{actions} are enclosed by braces. For example, the \texttt{constr} constructor defines two rules, the \textit{method} \texttt{met} and the \textit{always} \texttt{r0}, and the \texttt{prec!} statement specifies the \textit{precedence} relation between them, \texttt{met}$\prec$\texttt{r0}.} \frontend combines the benefits of traditional hardware transactions with flexible and parameterized construction support of Rust embedding.




\subsection{Temporal Hardware Transactions}
\label{sec:temp-ht}

We introduce the core features of \textit{temporal hardware transactions}: \textit {temporal relationships} and \textit{multi-cycle rules}. They provide a temporal view of hardware behavior and enable inter-cycle analysis. Both features are implemented in \frontend.

\subsubsection{Temporal Relationships.}
\label{sec:temp-relationship}

We define \textit{temporal relationships} among rules as two-fold: (1) \textit{temporal guard} specifies the fire condition of a rule based on the execution history of predecessor rules; and (2) \textit{temporal message passing} delivers data between temporally-related rules through \textit{channels} to live across clock cycles. 

{\compactparagraph{Syntax}
\autoref{fig:syntax}b shows the syntax of the \textit{temporal relationship} extension in \frontend. Specifically, for the current rule \texttt{r\_cur}, no matter \textit{method} or \textit{always}, it is guarded by a predecessor rule's execution history and communicates with the predecessor rule across cycles. It comprises four parts: (1) \textit{predecessor declaration}: alias the predecessor rule (\texttt{r\_pre}) to a local identifier (\texttt{p}); (2) \textit{channel declaration}: specifies the output channels (\texttt{o\_ch}); (3) \textit{temporal guard}: specifies the predecessor's execution history as part of the current rule's guard through the \textcolor{purple}{\texttt{delay}}, \textcolor{purple}{\texttt{dyndelay}}, or \textcolor{purple}{\texttt{eagerdelay}} operators; (4) \textit{message passing}: calls the builtin \texttt{recv} and \texttt{send} methods of the predecessor rule's channel (\texttt{p.ch}) and the current rule's channel (\texttt{o\_ch}) to deliver messages.}

\begin{figure}[t]
  \centering
    \includegraphics[width=\linewidth]{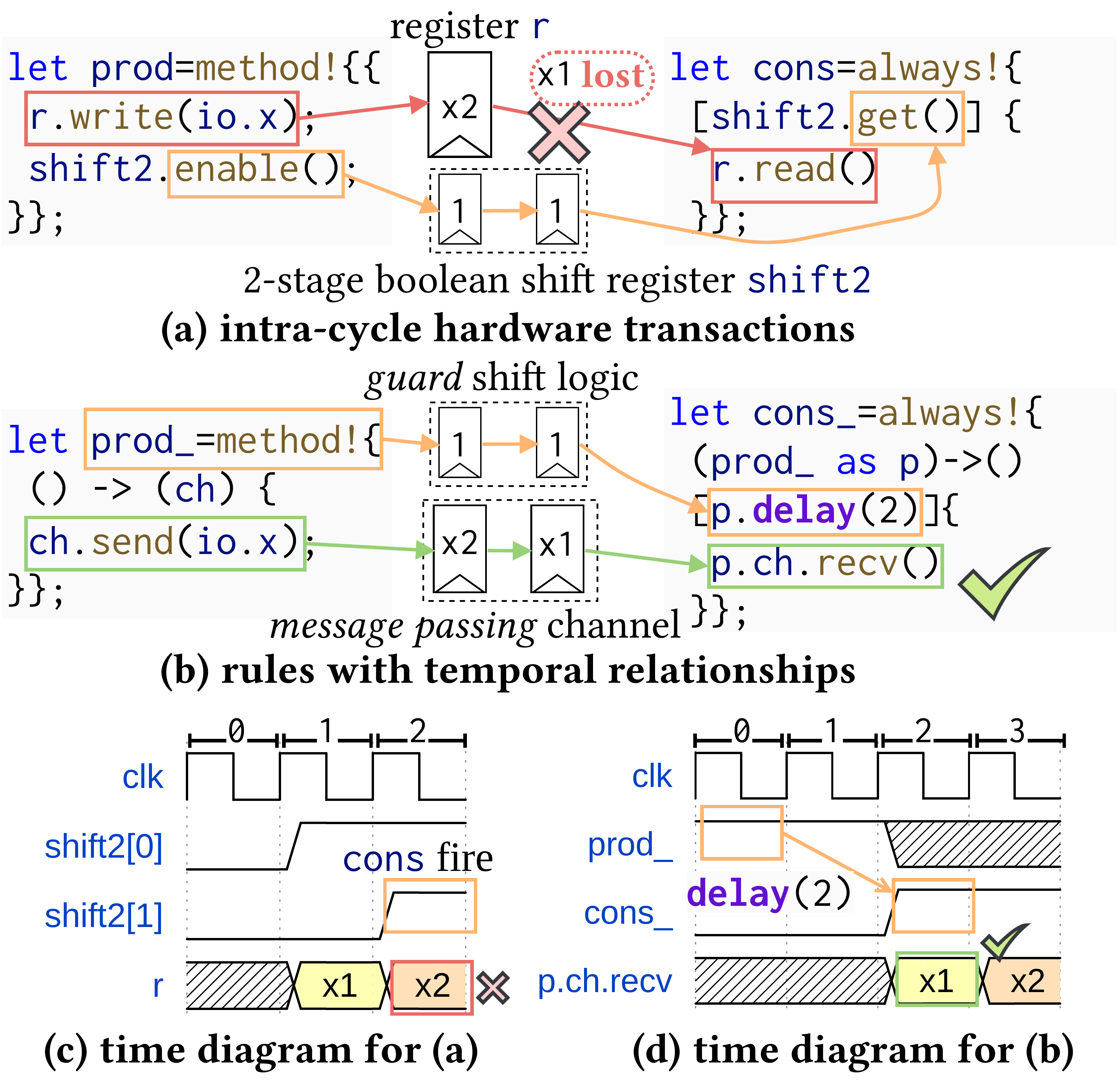}
    \caption{{Avoid producer-consumer mismatch.}}
    \label{fig:prod_cons_mismatch}
  \end{figure}
\begin{figure*}[t]
  \centering
  \includegraphics[width=.9\textwidth]{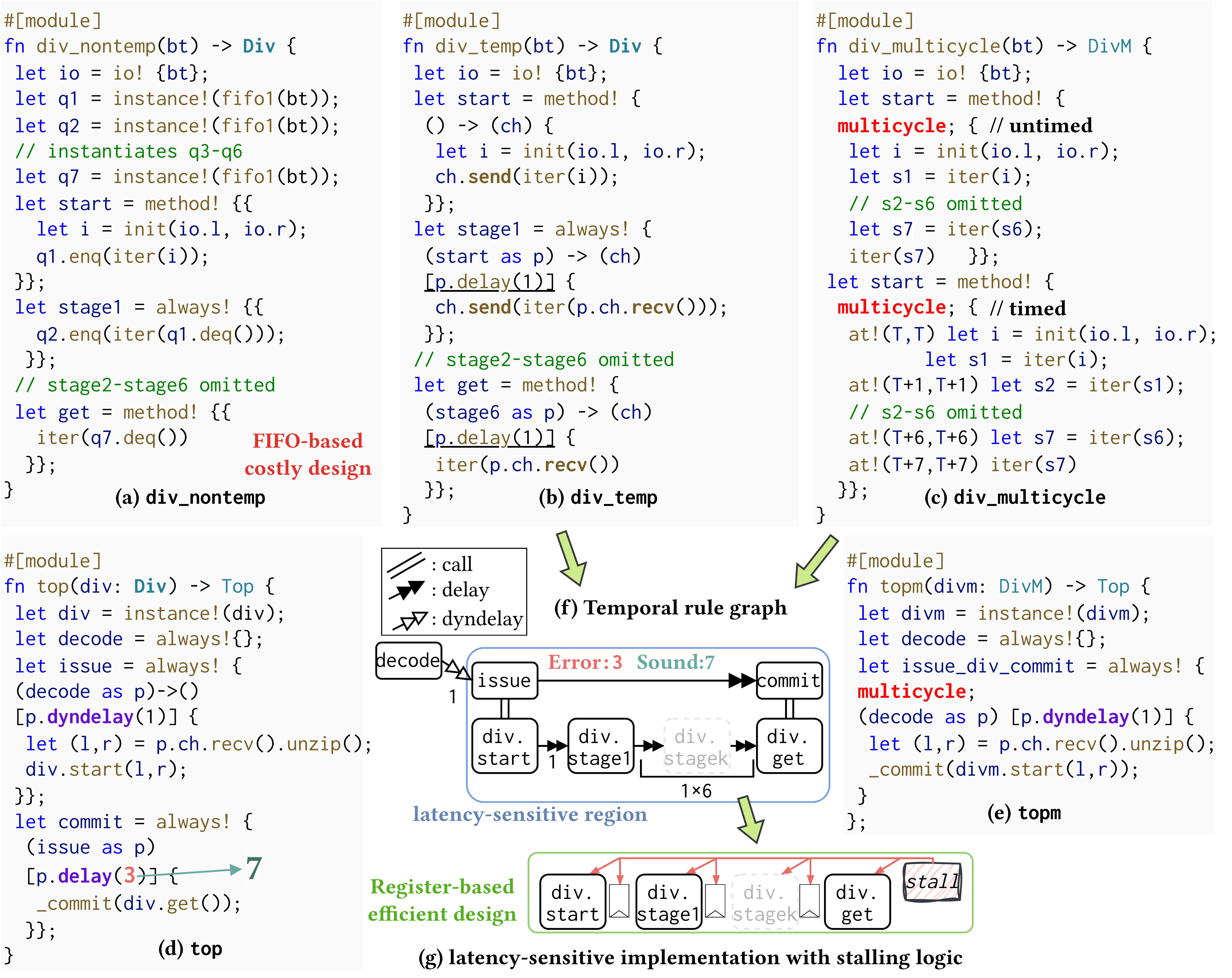}
  \caption{{Implementations of a 8-bit restoring division pipeline with \textit{temporal hardware transactions}.}}
  \label{fig:divider}
\end{figure*}

\compactparagraph{Semantics} 
The \textcolor{purple}{\texttt{delay}} operator specifies an interval of the fixed number of clock cycles as one guard condition, named \textit{latency-sensitive guard}. For example, \texttt{p.delay(k)} will hold exactly at clock cycle \texttt{T+k} if the predecessor rule \texttt{p} fires at clock cycle \texttt{T}. If the current rule cannot fire at clock cycle \texttt{T+k} due to other guard conditions' failure, the latency-sensitive guard will \textit{expire} and will not hold in subsequent clock cycles. The \textcolor{purple}{\texttt{dyndelay}} operator, on the other hand, specifies a variant interval of a minimum number of clock cycles as one guard condition, named \textit{latency-insensitive guard}. That is, \texttt{p.dyndelay(k)}'s holding time will start from clock cycle \texttt{T+k} until a successful firing of the current rule, if the predecessor rule \texttt{p} fires at clock cycle \texttt{T}. {The \textcolor{purple}{\texttt{eagerdelay}} operator is a variant of \textcolor{purple}{\texttt{delay}}, and is used only when the predecessor rule \texttt{p} is a \textit{multi-cycle rule} (\autoref{sec:m-rule}), and the delay countdown starts when \texttt{p} starts firing.} Temporal relationships' semantics require that \textit{temporal guards} must be coordinated with the \textit{channel-based message passing}, named \textbf{\textit{guard-message atomicity}}. Specifically, when a latency-sensitive guard expires, its carried message will be discarded from the channel at the same clock cycle; for a latency-insensitive guard, a message will remain in the channel until the guarded rule successfully fires. This temporal property helps avoid inter-cycle producer-consumer mismatch bugs: \textit{the produced data does not last long enough for the consumer to consume or does not arrive when the consumer is ready to consume}. \autoref{fig:prod_cons_mismatch} shows an example. \autoref{fig:prod_cons_mismatch}a uses intra-cycle hardware transactions: the producer rule \texttt{prod} writes the data \texttt{io.x} to the register \texttt{r}, and it calls \texttt{shift2.enable} to set the first stage of the 2-stage boolean shift registers \texttt{shift2} high. Its time diagram is shown in \autoref{fig:prod_cons_mismatch}c. The consumer rule \texttt{cons} is fired two cycles after the producer rule \texttt{prod}. However, it cannot read the data \texttt{x1}, which has been overwritten by the data \texttt{x2}, causing a \textit{data loss} bug. Instead, \autoref{fig:prod_cons_mismatch}b adopts the temporal relationship: the producer rule \texttt{prod\_} sends \texttt{io.x} to its output channel \texttt{ch}, and the consumer rule \texttt{cons\_} uses \texttt{prod\_} as the predecessor rule \texttt{p} and is fired two cycles after \texttt{prod\_} with \texttt{p.delay(2)} as guard, as shown by the time diagram in \autoref{fig:prod_cons_mismatch}d. The \textit{guard-message atomicity} guarantees that when \texttt{cons\_} is fired at cycle 2, it can receive the data \texttt{x1} coordinated with the rule firing, avoiding data loss.

\compactparagraph{Hardware implementation}
Both \textit{temporal guard} and \textit{temporal message passing} correspond to shifting logic: \textit{temporal guard} shifts the firing history of predecessor rules, and \textit{temporal message passing} shifts messages to keep pace with the guard. The \texttt{delay} is implemented as the efficient shift register for both guards and messages, while the \texttt{dyndelay} is implemented as a FIFO. 

\autoref{fig:divider} shows the description of an 8-bit restoring division pipeline in \frontend. Function \texttt{init} and \texttt{iter} include actions for the computation initialization and iteration, respectively, which can be considered as two combinational logic blocks. The (a)\texttt{div\_nontemp} constructor uses intra-cycle hardware transactions, describing the pipeline stages as rules (\texttt{start}, \texttt{stage1}, \ldots, \texttt{get}), each of which manipulates manually instantiated FIFO instances (\texttt{q1}, \ldots, \texttt{q7}) to coordinate one-by-one rule execution and deliver results. The (b)\texttt{div\_temp} constructor adopts \textit{temporal relationships} to describe the pipeline more intuitively with \texttt{p.delay(1)} temporal guards, and it uses channels to deliver results. 

\subsubsection{Multi-Cycle Rules.}
\label{sec:m-rule}

Although \textit{temporal relationships} opens the door for temporal behavior modeling, it has two-fold drawbacks. First, it still requires designers to describe multiple rules and specify temporal relationships among them manually, which is verbose. As exemplified in \autoref{fig:divider}, the (b)\texttt{div\_temp} constructor cannot reduce the number of lines of description code. Second, intra-cycle rules cannot be adjusted by the compiler, such as splitting a rule with a long critical path into multiple for frequency improvement. To tackle these problems, the \textit{temporal hardware transaction} abstraction further introduces \textit{multi-cycle rules}.

{
\compactparagraph{Syntax}
\autoref{fig:syntax}c shows the syntax of the \textit{multi-cycle rule} extension in \frontend. A \textit{multi-cycle rule} definition is distinguished by the \texttt{multicycle} keyword. The firing actions of a multi-cycle rule can be specified exactly the same as the intra-cycle rules. We provide the \textit{timing label} mechanism to specify the firing time (start time, finish time, or both) of the subsequent actions by the \texttt{at!} statements. Every timing label is associated with a \textit{timing variable} (e.g., \texttt{T} and \texttt{G}) and an optional constant offset (e.g., \texttt{1} in \texttt{T+1}).
}


\compactparagraph{Semantics}
When one multi-cycle rule is fired, its firing actions will be executed exactly \textit{once} in the subsequent clock cycles, named \textbf{\textit{multi-cycle atomicity}}. That is, when one action cannot execute in the current cycle due to \textit{dependency violations}, guard failures, or \textit{precedence violations}, the action will be retried in the subsequent cycles until it successfully executes. Action execution in multi-cycle rules must observe: (a) \textit{data dependency}, one action cannot be fired until all the input values are valid; (b) \textit{physical-timing dependency}, one action path of total delay exceeding the target clock period must be fired in different cycles.
{
In a timing label \texttt{T+k}, the timing variable \texttt{T} represents a certain clock cycle, and the constant \texttt{k} represents a latency-sensitive offset. \texttt{T+k} indicates the timing point \texttt{k} cycles after cycle \texttt{T}. For an action whose timing is specified by \texttt{at!(T+1,G)}, the action will start firing one cycle after \texttt{T} and finish firing at the clock cycle \texttt{G}. Different timing variables (\texttt{T} and \texttt{G}) indicate the unresolved latency, modeling the latency-insensitive temporal behavior.
} 

A multi-cycle rule can be either \emph{timed}, with a determined and legal \textit{schedule}, or \emph{untimed}, with unspecified action execution timing. Timed multi-cycle rules give designers precise cycle-level control over action execution, while untimed multi-cycle rules relieve designers from scheduling, leaving the tasks to the \textit{temporal scheduling} algorithm in \autoref{sec:t-sched}. Untimed multi-cycle rules are \textit{reusable}: they can have different \textit{schedules} for different configurations (e.g., target technology and frequency).  In \autoref{fig:divider}, the (c)\texttt{div\_multicycle} constructor describes the division pipeline by either an untimed or a timed multi-cycle rule. For the timed multi-cycle rule, all timing labels have the same timing variable \texttt{T} plus constant offsets, indicating \textit{latency-sensitive} behavior.

{
\subsubsection{Inter-cycle analysis and optimizations.}
\label{sec:inter-cycle-analysis}
Temporal hardware transactions boost productivity with intuitive syntax, and improve design robustness with semantic guarantees, including \textit{guard-message atomicity} and \textit{multi-cycle atomicity}. They also enable inter-cycle analysis and optimizations for compiler-enforced correctness and efficiency, beyond the scope of HDLs due to temporal unawareness.
}

\compactparagraph{Temporal rule graph}
A temporal hardware transaction module with temporal relationships and \emph{timed} multi-cycle rules can be abstracted as a unified \textit{temporal rule graph} representation for analysis, checking, and optimization, where vertices represent \textit{intra-cycle} rules and three types of edges represent different relationships among them: \textit{call}, \textit{delay}, and \textit{dyndelay}. For example, \autoref{fig:divider}d presents the \textit{temporal rule graph} for both the module (d)\texttt{top} instantiating (b)\texttt{div\_temp} and the module (e)\texttt{topm} instantiating (c)\texttt{div\_multicycle}. A \textit{latency-sensitive region} is defined as a \textit{connected component} of the \textit{temporal rule graph} with only \textit{call} and \textit{delay} edges remaining. In \autoref{fig:divider}d, \texttt{issue}, \texttt{commit}, and the divider rules form a latency-sensitive region.

\compactparagraph{Timing inference and rule coordination checking}
The compiler conducts \textit{timing inference} for each latency-sensitive region in a bottom-up manner. In \autoref{fig:divider}, \texttt{commit} can be inferred to fire 7 clock cycles after \texttt{issue}: \texttt{commit} fires together with the callee \texttt{div.get}, which fires 7 cycles after \texttt{issue}. {The compiler leverages the inferred timing information for \textit{rule coordination checking}. In \autoref{fig:divider}d, \texttt{commit} has the \textit{temporal guard} \texttt{issue.delay(3)}, which mismatches the inferred delay 7. This mis-coordination causes the \texttt{commit} rule never to fire since the temporal guard holds when \texttt{div.get} cannot be called. This causes \textit{deadlock}s in practice. For example, permanent commit failure in a CPU pipeline makes the scoreboard full and stalls the pipeline forever. The \textit{rule coordination checking} reports such mis-coordination as a \textit{temporal error} and provides a fix suggestion with the inferred timing. }

\compactparagraph{Temporal relationship pruning}
We define \textit{redundant} temporal relationships as the ones implied by either \textit{guarded atomicity} of its target rule or other temporal relationships ending at the same target rule. The compiler conducts \textit{temporal relationship pruning} together with the bottom-up \textit{timing inference}. When the timing of a rule is inferred, the pruning process removes all the temporal relationships targeting the rule if the timing is implied by its callees, or only keeps the one of the shortest delay. From the perspective of the \textit{temporal rule graph}, the pruning transforms each \textit{latency-sensitive region} into a \textit{tree}, keeping the temporal behavior unchanged with implementation simplified. It is worth noting that the \textit{temporal relationship pruning} only removes \textit{temporal guards}, with all \textit{temporal message passing channels} remaining.

{
\compactparagraph{False static prevention}
A \textit{false-static} pattern denotes that a rule has one \textit{delay} guard accompanied by other guards after \emph{temporal relationship pruning}. When the \textit{delay} guard holds but any other guard fails, the rule cannot fire, and the latency-sensitive guard will expire immediately with the message discarded, causing \textbf{data loss} bugs. The compiler detects and reports \textit{warning}s on the \textit{false-static} pattern, recommending \textit{dyndelay} operators for robustness.
}

\subsubsection{Hybrid latency-sensitive/-insensitive temporal behavior support.}
\label{sec:hybrid-m-rule}
Both \textit{temporal relationships} and \textit{multi-cycle rules} can describe hybrid latency-sensitive/-insensitive temporal behavior. {Any hybrid temporal behavior can be represented as \textit{latency-sensitive regions} connected by \textit{dyndelay} edges in the \textit{temporal rule graph}. \autoref{fig:divider}f presents one example, where \texttt{decode} can be treated as a latency-sensitive region of a single rule, and it is connected, with a \textit{dyndelay} edge, to the latency-insensitive region including the \texttt{issue} rule. The compiler automatically implements the hybrid behavior by adding minimal \textit{stall} logic, wrapping each latency-sensitive region with only one \textit{stall} controller, which controls the state updating of the whole region, including both the rules and the temporal relationship implementations, as shown in \autoref{fig:divider}g. The \textit{stall} controller watches all the \textit{dyndelay} edges connected to the region. When any of them is blocked, either being full for sending or empty for receiving, the \textit{stalling} logic will stall the whole region, preventing wrong behavior such as data loss or using invalid data. }



\section{Compiler}
\label{sec:compiler}
\begin{figure}[t]
  \centering
  \includegraphics[width=\linewidth]{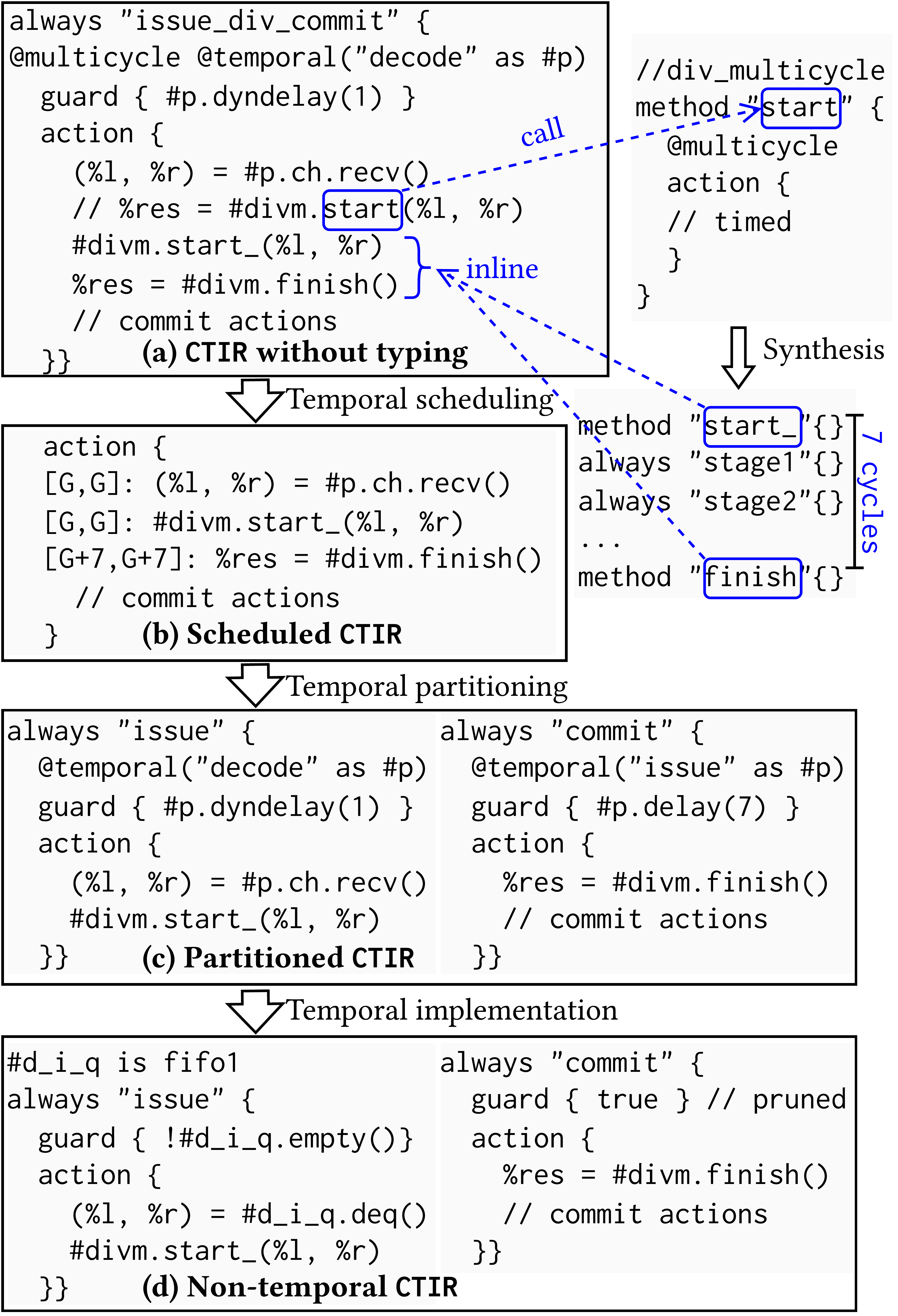}
  \caption{{\ctir and synthesis flow}}
  \label{fig:ir-trans}
\end{figure}

\cmt's compiler is built around the \cmt Transaction Intermediate Representation (\ctir) and conducts multi-phase synthesis to generate rich backends. \ctir provides a unified representation for both \textit{temporal} and intra-cycle hardware transaction abstraction. \autoref{fig:ir-trans}a shows the \ctir representation of the untimed multi-cycle rule \texttt{topm} in \autoref{fig:divider}e. It shows that \ctir strictly aligns with the language features of \frontend, facilitating straightforward IR construction. For example, \textit{timing labels} in multi-cycle rules appear in \ctir, as shown in \autoref{fig:ir-trans}b, instructing the compiler to generate expected schedules and implementations.


The compiler translates \ctir of \textit{temporal hardware transaction} features into a low-level description of the same functionality. 
The synthesizer works in a \textit{bottom-up} manner: any callee multi-cycle rules should be synthesized before the current rule. When synthesizing a multi-cycle rule, the compiler will first \textit{inline} all method calls to other multi-cycle rules. It will substitute the original method call with a call to the entry method among the synthesized intra-cycle rules. Besides, the compiler also inserts method calls to get the result values. For example, in \autoref{fig:ir-trans}a, the method call to \texttt{\#divm.start} is replaced by the partitioned \texttt{\#divm.start\_}, and a call to \texttt{\#divm.finish} is added to retrieve the result \texttt{\%res}. The synthesis process comprises the following phases:

\compactparagraph{Temporal scheduling}
\label{sec:t-sched}
The \textit{temporal scheduling} phase generates legal schedules for untimed multi-cycle rules. Since \textit{temporal hardware transactions} support hybrid latency-sensitive/-insensitive temporal behavior as discussed in \autoref{sec:hybrid-m-rule}, we cannot adopt existing scheduling algorithms, such as SDC~\citep{cong_efficient_2006}, since they only assign operations into \textit{static} timing positions. Instead, we introduce a custom ASAP (As-Soon-As-Possible) scheduling algorithm. The scheduler iterates the actions of the current rule and determines their earliest firing timing in order. For example, the scheduler groups the action \texttt{\#p.ch.recv} and \texttt{\#divm.start\_} to be fired at clock cycle \texttt{G}, assuming the \textit{physical-timing dependency} between them is satisfied. Next, the scheduler puts the action \texttt{\#divm.finish} into the cycle \texttt{G+7} according to the timing reported from \texttt{\#divm}. The remaining commit actions consuming \texttt{\%res} must be scheduled not earlier than \texttt{G+7} to satisfy the \textit{data dependency}. For hybrid latency-sensitive/-insensitive temporal behavior, our ASAP scheduler will automatically create new timing variables for actions whose firing time cannot be determined statically. For example, if \texttt{\#divm} is a latency-insensitive division unit, the firing time of the action \texttt{\#divm.finish} cannot be assigned a timing label in the \texttt{G+k} form. Instead, the scheduler will create a new timing variable (e.g., \texttt{T}) to indicate the firing time. The remaining commit actions will be scheduled at later cycles than \texttt{T}.

Our scheduler performs \emph{retiming} on multi-cycle rules. Specifically, given a technology-specific propagation delay model and a clock period target, the scheduler considers the \textit{physical-timing dependencies} for each action: it iterates the scheduled actions that start a data dependency path to the current action, and use the propogation delay model to estimate the path delay to determine the earliest clock cycle that the current action can be scheduled.

\compactparagraph{Temporal partitioning}
\label{sec:t-partition}
The \textit{temporal partitioning} pass partitions a timed multi-cycle rule into an equivalent set of intra-cycle rules with \textit{temporal relationships} generated. It involves three steps: (1) building a new rule for every group of actions that have the same timing label; (2) creating \textit{temporal guards} between new rules according to the schedule; (3) inserting \textit{temporal message passing} channels and actions according to \textit{data dependency}. For example, in \autoref{fig:ir-trans}b, the partitioning pass creates two new rules \texttt{issue} and \texttt{commit} from the scheduled groups, and inserts a \textit{latency-sensitive guard} of the 7-cycle delay between them, generating \autoref{fig:ir-trans}c.
This example does not create message passing channels since there is no data dependency across rule boundaries. {For any dependency between two rules whose timing labels have different timing variables, the compiler will create a \textit{latency-insensitive guard} to enforce the dependency and a channel for data delivery if required, which guarantees \textit{multi-cycle atomicity}. This phase only creates necessary latency-insensitive logic, keeping the implementation efficient.}

\compactparagraph{Temporal implementation}
The \textit{temporal implementation} pass translates \textit{temporal guards} and \textit{message passing} channels into non-temporal instances and actions. Before the implementation, the compiler builds the \textit{temporal rule graph} representation and conducts \textit{inter-cycle} analysis, checking, and optimizations, as described in \autoref{sec:inter-cycle-analysis}. In \autoref{fig:ir-trans}d, \texttt{d\_i\_q} is instantiated to implement the temporal guard and message passing between rule \texttt{issue} and its predecessor \texttt{decode}. The temporal relationship between \texttt{issue} and \texttt{commit} is optimized out by \textit{temporal relationship pruning}.


The \cmt compiler supports rich backends. For RTL generation, it implements an efficient hardware transaction synthesis algorithm~\citep{bluespec_inc_system_2025,noauthor_b-lang-orgbsc_2025}. Specifically, it translates \ctir into FIRRTL~\citep{izraelevitz_reusability_2017}, and generates optimized SystemVerilog using firtool-1.108.0~\citep{circt_community_release_2025} for Vivado synthesis and FPGA deployment. \frontend supports transactional testbenches: test stimuli are programmed as \textit{rules}, which peek and poke instances through method calls. \cmt supports the RTL simulator generator Verilator~\citep{veripool_verilator_2025} and generates C++ harness code from testbenches to drive the RTL simulators.

\section{Evaluation}

\label{sec:evaluation}

We evaluate \cmt from four perspectives, including a general soft processor, custom instructions, linear algebra accelerators, and systolic array design. We discuss how temporal hardware transactions facilitate the case studies, as summarized in \autoref{tab:impl}, and analyze design quality for FPGA implementation.

\begin{table}[t]
    \centering
    \small
    \setlength{\tabcolsep}{1pt}
    \caption{{Temporal features in evaluation designs}}
    \label{tab:impl}
    \begin{tabular}{|c|c|c|c|c|}
    \hline
    Design & \begin{tabular}[c]{c}Temporal\\ relationship\end{tabular} & \begin{tabular}[c]{c}Multi-cycle\\ rule\end{tabular}   & SLOC &\begin{tabular}[c]{c}Synth.\\ interface\end{tabular}  \\
    \hline
    Soft processor & \dcheck & untimed & 571 & LI \\
    Rgba2gray, sobel, .. & \dcheck & untimed & 86 & Hybrid \\
    Polybench kernels& \dcheck & untimed & 771 & LS \\
    Systolic array & \dcheck & timed & 35 & LS \\
    \hline 
    \end{tabular}
\end{table}



\subsection{5-stage RISC-V Soft Processor}
\label{sec:cpu-core}

\newcommand{\cmtRV}{\texttt{CMT2-RV}\xspace}

\begin{table}[t]
  \centering
  \small
  \setlength{\tabcolsep}{3pt}
  \caption{{Evaluation results of RISC-V soft cores.}}
  \label{tab:cpu-base-results}
  \begin{tabular}{|c|c|c|c||c|}
    \hline
     & Sodor~\citep{noauthor_ucb-barriscv-sodor_2025} & HF~\citep{jang_modular_2024}  & \cmtRV & \makecell[c]{\cmtRV + \\ Rgba2gray, sobel, ..} \\
    \hline
    CPI & 1.389 & 1.389 & 1.386 & -  \\
    Frequency & 367MHz & 287MHz & 377MHz & 316MHz \\
    LUT & 1974 & 3055 & 1614 & 2729 \\
    FF & 924 & 2829 & 779 & 1152 \\
    \hline
  \end{tabular}
\end{table}


\compactparagraph{Implementation}
We implement a RISC-V soft processor, denoted as \cmtRV, in \frontend according to the architecture of Sodor~\citep{noauthor_ucb-barriscv-sodor_2025} (5-stage, fully bypassed). Our design adopts \textit{temporal hardware transactions}, as illustrated in \autoref{fig:cpu-temporal}, especially {using temporal relationships between pipeline stages and an \textit{untimed} multi-cycle rule to describe backend behavior shown in \autoref{fig:cpu-temporal}. The multi-cycle rule is scheduled into latency-insensitive stages that support stalling for hazard resolution.}

{
\compactparagraph{Baselines}
We compare \cmtRV with two baselines: Sodor in Chisel~\citep{bachrach_chisel_2012} and Sodor in HazardFlow~\citep{jang_modular_2024}. Chisel is an embedded HDL for RTL design, and HazardFlow is an academic HDL featuring latency-insensitive pipeline description. Other approaches, including HLS~\citep{amd_inc_vitis_2025,josipovic_dynamically_2018}, Calyx~\citep{nigam_compiler_2021,kim_unifying_2023}, and Filament~\citep{nigam_modular_2023}, are not compared since they cannot describe the required architecture.}

\compactparagraph{Methodology}
To measure cycles per instruction (CPI), we use MachSuite~\citep{reagen_machsuite_2014} integer benchmarks and EEMBC CoreMark~\citep{noauthor_eembccoremark_2025} benchmarks and run RTL simulation to collect the cycle counts. We use Verilator~\citep{veripool_verilator_2025} v5.028 to simulate \cmtRV. We synthesize and place-and-route all the cores with Vivado 2024.1, targeting an XCVU9P FPGA. We exclude memories for all the designs.

\compactparagraph{Results}
RTL simulation shows that all five cores achieve almost the same CPI. It indicates that \cmt's \textit{temporal hardware transactions} have the expressiveness to describe the required processor features, including data forwarding and branch prediction. \autoref{tab:cpu-base-results} shows that \cmtRV has the frequency of 377MHz, higher than Sodor's 367MHz. The HazardFlow core only achieves 287MHz. For resource usage, HazardFlow uses 1.55\x LUTs and 3.06\x FFs compared to Sodor, which is a huge overhead. Instead, \cmtRV uses 0.82\x LUTs and 0.84\x FFs compared to Sodor, indicating efficient resource usage. These results demonstrate that \cmt's high-level abstraction does not sacrifice design quality for general hardware.

\subsection{Custom CPU Instructions}
With \textit{temporal hardware transactions}, we can easily extend custom instructions to the \cmtRV core in \autoref{sec:cpu-core}. By reserving an extension interface in \cmtRV's constructor, we can add new instructions by providing instruction encoding and rules for behavior description. We evaluate an image processing workload, \textit{Edge Detection}, using the same methodology as \autoref{sec:cpu-core} for evaluation. This case study aims to demonstrate that \cmt provides a convenient and efficient way to model and evaluate architecture decisions like adding instructions.

{
\compactparagraph{Implementation} We accelerate \textit{Edge Detection} with custom instructions including \texttt{rgba2gray}, \texttt{sobel3x3}, \texttt{erode3x3}, and \texttt{dilate3x3}, all of which are described as untimed multi-cycle rules with the loop control described as temporal relationships according to software branching among basic blocks, as exemplified in \autoref{fig:nested-loop}. The synthesized accelerators include hybrid latency-sensitive/-insensitive implementation as introduced in \autoref{sec:hybrid-m-rule}, since their behavior contains latency-insensitive memory access. It only takes 86 SLOC in total for the behavior description of these custom instructions, presenting software-level productivity.
}

\compactparagraph{Results} With the four custom instructions, the cycle count, reported by RTL simulation, is reduced by 75\% compared to the original \texttt{Edge Detection} workload running without custom instructions. \autoref{tab:cpu-base-results} shows the frequency and resource usage results of the extended \cmtRV core. The synthesized frequency is 316MHz, which is 16\% lower than the original \cmtRV core. For resource usage, the extended \cmtRV core uses 1.69\x LUTs and 1.48\x FFs compared to the original one, reporting the real hardware overheads. \cmt reports comprehensive performance results, including cycle count and frequency, and real resource usage, given the high-level description of the compound system, including both a processor and custom instructions. It saves human efforts to program at a tedious and error-prone low level to get the precise evaluation results.

\subsection{Linear Algebra Kernels}
\label{sec:polybench}

\begin{figure}[t]
  \centering
  \includegraphics[width=.95\linewidth]{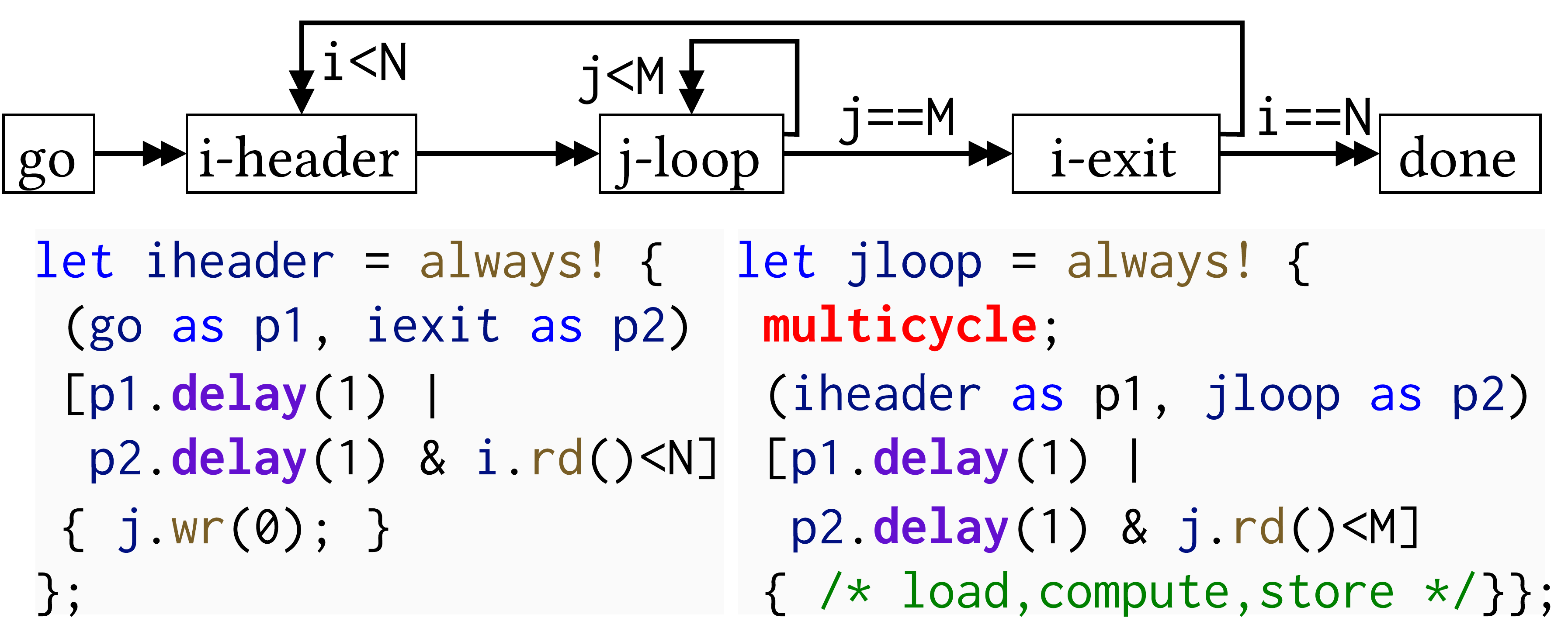}
  \caption{Nested loops in \cmt}
  \label{fig:nested-loop}
\end{figure}

We evaluate linear algebra kernels from PolyBench~\citep{louis-noel_pouchet_polybenchc_2018} to demonstrate \cmt's capability for control-intensive designs.

\compactparagraph{Implementation}
We implement resource-efficient accelerators for PolyBench kernels, which contain nested loops to be described as hardware finite state machines (FSMs). \cmt's abstraction facilitates such multi-cycle behavior description, as illustrated in \autoref{fig:nested-loop}. {The description concisely corresponds to the control flow graph (CFG) of the kernel, where transitions between basic blocks are described as temporal relationships, and complex basic blocks are described as multi-cycle rules, such as \texttt{jloop}. We implement 13 kernels from the benchmark suite, and the total SLOC is 771, including 121 rules (21 of them are multi-cycle rules, all untimed). All the kernels are latency-sensitive, since on-chip RAM access has a fixed latency.} It only takes a Ph.D. student one day to implement. The same kernels in SystemVerilog~\citep{design_automation_standards_committee_ieee_2024} take 2610 SLOC in total, and it takes a Ph.D. student 1 week to implement and test. 

\compactparagraph{Baselines}
{We compare \cmt-synthesized designs with five baselines: SystemVerilog for manual RTL design, BSV~\citep{nikhil_bluespec_2004} for traditional rule-based design, Cement~\citep{xiao_cement_2024} and Dahlia-Calyx~\citep{nigam_predictable_2020,nigam_compiler_2021} flow for automatic FSM generation from software-like description, and Vitis HLS 2024.1~\citep{amd_inc_vitis_2025} considered as the commercial state-of-the-art HLS tool. BSV generates FSMs by \texttt{Stmt}~\citep{bluespec_inc_bluespec_nodate}. Static optimizations~\citep{kim_unifying_2023} are enabled in the Dahlia-Calyx flow. Filament~\citep{nigam_modular_2023} is not included since it cannot describe loop control, while dynamic HLS approaches~\citep{josipovic_dynamically_2018,xu_eliminating_2023,cheng_combining_2020} are also excluded since they present disadvantages against Vitis HLS on static kernels.
All designs are configured to be sequential and use synchronous-read RAM for comparison fairness. }
We use {Vivado 2024.1} for synthesis and place-and-route toward the XCVU9P FPGA with a target period of 7ns. 


\begin{table}[t]
  \centering
  \small
  \setlength{\tabcolsep}{1.5pt}
  \caption{{Performance and resource across Polybench kernels.}}
  \label{tab:polybench-results}
  \begin{tabular}{|c|c|c|c|c|c|c|}
    \hline
     & SV & BSV\citep{nikhil_bluespec_2004} & Cement\citep{xiao_cement_2024} & Vitis HLS\citep{amd_inc_vitis_2025} & Calyx\citep{nigam_compiler_2021} & \cmt \\
    \hline
    Cycle & \textit{1204} & 1.59\texttimes & 0.92\texttimes & 1.01\texttimes & 3.83\texttimes & 0.92\texttimes \\
    Time  & \textit{5.5\textmu s} & 1.74\texttimes & 1.14\texttimes & 1.48\texttimes & 2.73\texttimes & 1.01\texttimes \\
    LUT   & \textit{431} & 1.87\texttimes & 2.01\texttimes & 1.33\texttimes & 1.19\texttimes & 0.87\texttimes \\
    FF    & \textit{169} & 3.06\texttimes & 0.73\texttimes & 2.23\texttimes & 2.12\texttimes & 0.81\texttimes \\
    \hline
  \end{tabular}
\end{table}
\compactparagraph{Results}
\autoref{tab:polybench-results} shows the geometric mean performance and resource utilization results across the PolyBench kernels. For performance, \cmt achieves better cycle counts (0.92\texttimes\xspace) and comparable execution time (1.01\texttimes\xspace) than the SystemVerilog baseline. {The reason is that \cmt designs avoid redundant cycles compared to the manually crafted FSMs with acceptable frequency overhead.} Although Cement and Vitis HLS achieve similar cycle counts as \cmt, \cmt designs achieve higher frequencies and faster execution time. The performance of BSV and Dahlia-Calyx designs is not competitive since their frontends introduce unnecessary idle cycles for loop control and computation. For resource usage, \cmt saves 13\% LUTs and 19\% registers than the SystemVerilog baseline. {By observing Vivado-reported schematics, we found that \cmt compiler generates FSMs of one-hot encoded states natively from \textit{temporal guards}, which is friendly for FPGA synthesis and can save FSM resources.} Cement uses fewer registers than the SystemVerilog baseline but consumes the most LUTs due to extremely compact states and complicated transitions in the generated FSMs. Other approaches require more LUTs and registers than the SystemVerilog baseline. For example, BSV generates 24 rules for state transition, 42 rules for memory access, and 9 rules for FIFOs to implement the simple \texttt{atax} kernel. {Most of the verbose rules handle latency-insensitive actions. \cmt avoids such overheads since its efficient synthesis flow does not introduce unnecessary latency-insensitive logic.} The results demonstrate that \cmt boosts productivity while achieving competitive performance against handcrafted RTL implementation.

\subsection{Systolic Array Accelerators}
\label{sec:systolic-array}

We evaluate \cmt's applicability to high-performance architecture for FPGA acceleration. 

\begin{figure}[t]
  \begin{minipage}[b]{.55\linewidth}
    \begin{subfigure}[b]{\linewidth}
      \centering
      \includegraphics[width=.7\linewidth]{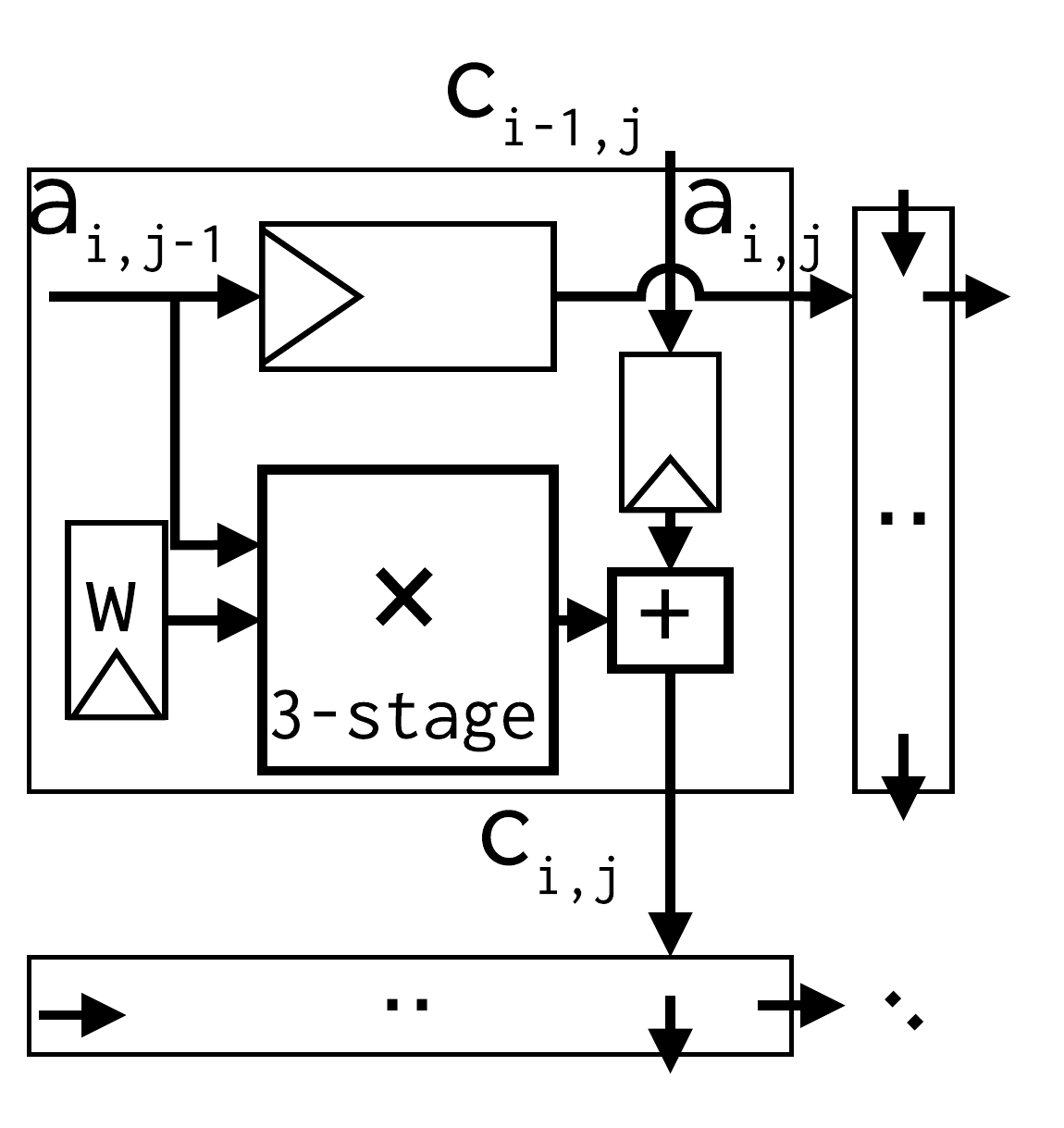}
      \caption{PE and interconnects}
      \label{fig:sa-pe}
    \end{subfigure}
    \begin{subfigure}[b]{\linewidth}
      \includegraphics[width=\linewidth]{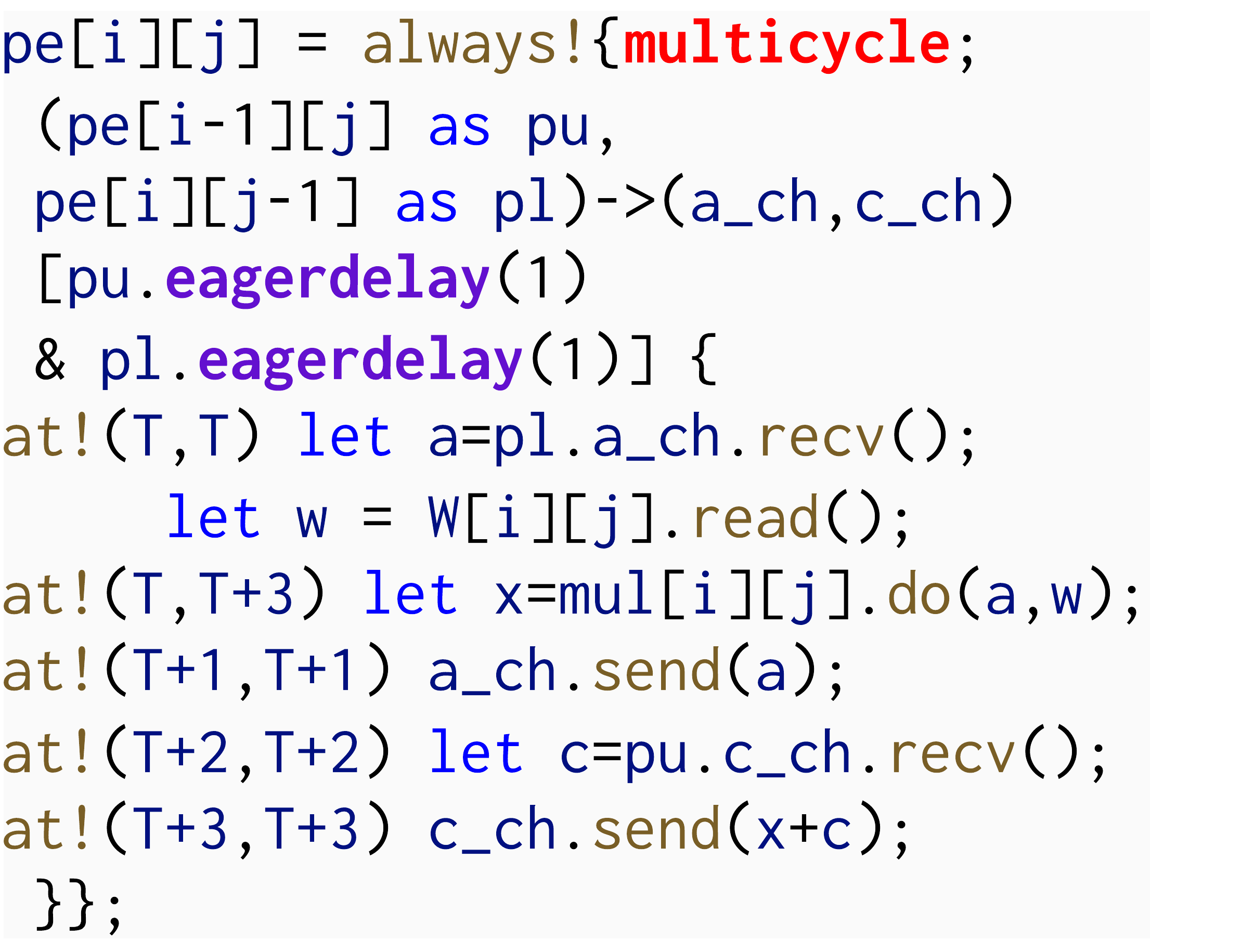}
      \caption{\cmt implementation}
      \label{fig:sa-pe-clamp}
    \end{subfigure}
  \end{minipage}
  \begin{minipage}[b]{.44\linewidth}
    \begin{subfigure}[b]{\linewidth}
      \centering
      \includegraphics[width=\linewidth]{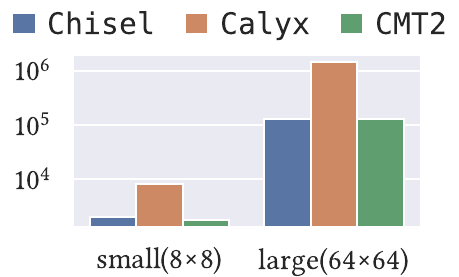}
      \caption{LUTs}
      \includegraphics[width=\linewidth]{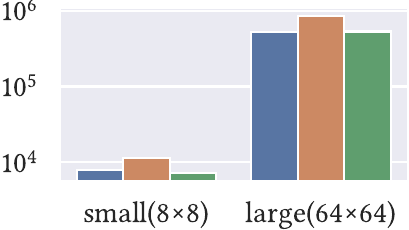}
      \caption{Registers}
      \includegraphics[width=\linewidth]{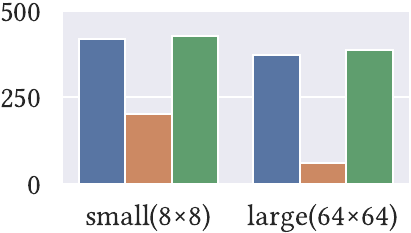}
      \caption{Frequency(MHz)}
      \label{fig:systolic-array-results}
    \end{subfigure}
  \end{minipage}
  \caption{{Systolic array implementation and evaluation.}}
\end{figure}

\compactparagraph{Implementation}
We implement a weight-stationary systolic array, whose processing element (PE) and interconnects are illustrated in \autoref{fig:sa-pe}. Such spatial architecture exhibits temporal behavior across clock cycles: every PE receives data from neighboring PEs, computes or temporarily stores data, and then sends data to other PEs. Thereby, \cmt's temporal hardware transactions can be easily applied to describe the behavior of the systolic array. {As shown in \autoref{fig:sa-pe-clamp}, we describe the behavior of a PE as a \textit{timed} multi-cycle rule and use temporal relationships to describe interconnects. Every PE gets data from its up and left neighbor, named \texttt{pu} and \texttt{pl}, respectively. The \textit{temporal guard} of the current PE is set by two \texttt{eagerdelay} operators. The timing labels control the received data to be delayed by one cycle before being sent to channels, implementing systolic data movement. Besides, the timing labels also specify the pipeline depth of the multiplier. This demonstrates temporal hardware transactions' precise control over behavior to align with the architectural design intent.}
The complete temporal description of the systolic array passes the compiler checking in \autoref{sec:inter-cycle-analysis} to guarantee rule coordination and no data loss, and is synthesized into a fully latency-sensitive implementation.

\compactparagraph{Baselines}
We compare the \cmt-synthesized systolic array with a high-performance weight-stationary systolic array in Chisel~\citep{luo_rubick_2023} and an output-stationary design produced by the newest Calyx systolic array generator~\citep{kim_unifying_2023}. We run Vivado 2024.1 with a target period of 2.5ns for the XCU250 FPGA part. We target 32-bit fixed-point GEMM, and synthesize multipliers into DSP slices.

\compactparagraph{Results}
\autoref{fig:systolic-array-results} shows the resource and frequency results. \cmt designs save geomean 7\% LUTs and 4\% registers than Chisel designs, and save 86\% LUTs and 38\% registers than Calyx designs. The minor differences between the resource usages of the \cmt designs and those of the Chisel designs demonstrate that \cmt does not introduce unnecessary overhead than handcrafted designs. The significant resource savings of \cmt over Calyx are due to Calyx's static inference of the clock-cycle timing of all data transfers among PEs and its generation of centralized FSMs to control all their execution, which incurs large overheads and worse frequencies. \cmt designs achieve slightly higher frequencies, 1.03\texttimes\xspace, than those of Chisel. These results demonstrate \cmt's hardware quality for high-performance architecture.
\section{Related Work}
\label{sec:related-work}
\compactparagraph{Embedded HDLs} Embedded HDLs leverage software languages for metaprogramming~\citep{bachrach_chisel_2012, lockhart_pymtl_2014, clow_pythonic_2017,jane_street_hardcaml_2025,papon_spinalhdl_2025} and support flexible parameterization and construction. This line of work remains at the structural register-transfer level without raising design abstraction to provide better language promises about behavioral correctness. 


\compactparagraph{Transactional HDLs} Transactional HDLs~\citep{bourgeat_essence_2020,nikhil_bluespec_2004,choi_kami_2017} embody the \textit{guarded atomic action} concurrency model for digital hardware design. Synthesis algorithms~\citep{hoe_synthesis_2000,hoe_operation-centric_2004,bluespec_inc_system_2025} generate RTL circuits from rule descriptions. Prior rule-based language extensions~\citep{greaves_further_2019,karczmarek_synthesis_2008} generate additional hardware units like arbiters and reservation stations, causing unavoidable overheads. Although \cmt introduces high-level abstraction, the temporal features are synthesized into efficient low-level implementations without introducing any unnecessary hardware components.

\compactparagraph{Type systems and models for hardware} Filament~\citep{nigam_modular_2023} and Aetherling~\citep{durst_type-directed_2020} encode latency-sensitive timing properties in their type systems, while Shakeflow~\citep{han_shakeflow_2023} and HazardFlow~\citep{jang_modular_2024} introduce latency-insensitive combinator interfaces. Their type systems detect hardware issues such as resource conflicts and combinational loops. However, they are purpose-built for only latency-sensitive or latency-insensitive. Similarly, PDL~\citep{zagieboylo_pdl_2022}, Cement~\citep{xiao_cement_2024}, Spade~\citep{skarman_spade_2022}, {TL-Verilog~\citep{hoover_timing-abstract_2017}}, and Esterel~\citep{berry_esterel_1992} introduce language constructs to facilitate specific designs, such as control-intensive or pipeline circuits. \cmt's language features provide a general description of temporal behavior for hybrid latency-sensitive/-insensitive scenarios. Assassyn~\citep{weng_assassyn_2025} provides an asynchronous programming model for both architecture and hardware design. \cmt's abstraction reaches a higher level to describe multi-cycle behavior while providing rich temporal analysis and synthesis capabilities.

\compactparagraph{High-level synthesis and IRs} High-level synthesis (HLS)~\citep{google_inc_xls_2025,amd_inc_vitis_2025,zhang_autopilot_2008, canis_legup_2011, xu_hermes_2024,josipovic_dynamically_2018} starts with software programs and generates hardware implementations. They highly rely on synthesis algorithms like scheduling, which generate either static~\citep{cong_efficient_2006,zhang_sdc-based_2013}, dynamic~\citep{josipovic_dynamically_2018,xu_crush_2025,xu_suppressing_2024}, or hybrid~\citep{xu_hector_2022,xu_eliminating_2023,cheng_combining_2020} circuits for design tradeoffs. However, HLS's abstraction is too high to expose enough control over hardware details, causing expressiveness limitations~\citep{agarwal_comparative_2010} and unpredictable performance~\citep{nigam_predictable_2020}. Hardware IRs~\citep{nigam_compiler_2021,xu_hector_2022,majumder_hir_2024,kim_unifying_2023} provide a middleground where high-level features co-exist with structural hardware. \cmt's high-level abstraction provides productivity even close to software description for temporal behavior, but does not sacrifice expressiveness and low-level control.

\section{Conclusion and Future Work}

We present \cmt, a new FPGA programming approach with the raised abstraction, \textit{temporal hardware transactions}, to provide an expressive description of temporal behavior for boosted productivity. We conduct comprehensive case studies, including soft processors, custom instructions, linear algebra kernels, and a systolic array, to demonstrate the effectiveness. For future work, we will perform real-world tasks like design exploration of out-of-order processors and heterogeneous systems with \cmt's help.

\bibliographystyle{ACM-Reference-Format}
\bibliography{3-CMT2}

\end{document}